\documentclass[aip,rsi,reprint,graphicx]{revtex4-1}
\usepackage{color}
\usepackage{lineno}
\usepackage{graphicx}
\usepackage{booktabs}
\usepackage{array}
\usepackage{braket}
\usepackage{amsmath,bm}
\usepackage{ulem}
\usepackage[
            colorlinks,
            linkcolor=blue,
            anchorcolor=blue,
            citecolor=blue,
            urlcolor=blue
            ]{hyperref}
\draft 

\begin{document}


\title{A New Design of Resonant Cavity for the W-band EPR spectrometer} 



\author{Yu He}
\affiliation{CAS Key Laboratory of Microscale Magnetic Resonance and School of Physical Sciences, University of Science and Technology of China, Hefei 230026, China}
\affiliation{CAS Center for Excellence in Quantum Information and Quantum Physics, University of Science and Technology of China, Hefei 230026, China}

\author{Runqi Kang}
\affiliation{CAS Key Laboratory of Microscale Magnetic Resonance and School of Physical Sciences, University of Science and Technology of China, Hefei 230026, China}
\affiliation{CAS Center for Excellence in Quantum Information and Quantum Physics, University of Science and Technology of China, Hefei 230026, China}

\author{Zhifu Shi}
\affiliation{Chinainstru $\&$ Quantumtech (Hefei) Co.,Ltd, Hefei 230031, China}

\author{Xing Rong}
\email{xrong@ustc.edu.cn}
\affiliation{CAS Key Laboratory of Microscale Magnetic Resonance and School of Physical Sciences, University of Science and Technology of China, Hefei 230026, China}
\affiliation{CAS Center for Excellence in Quantum Information and Quantum Physics, University of Science and Technology of China, Hefei 230026, China}

\author{Jiangfeng Du}
\email{djf@ustc.edu.cn}
\affiliation{CAS Key Laboratory of Microscale Magnetic Resonance and School of Physical Sciences, University of Science and Technology of China, Hefei 230026, China}
\affiliation{CAS Center for Excellence in Quantum Information and Quantum Physics, University of Science and Technology of China, Hefei 230026, China}


\date{\today}

\begin{abstract}
We report a new design of resonant cavity for W-band EPR spectrometer.
It suits with both solenoid-type and split-pair magnets.
The cavity operates on the TE$_{011}$ mode, where the microwave magnetic field is along the cylindrical axis.
Its cylindrical axis is horizontal, so the magnetic field of the microwave is always perpendicular to the vertical external magnetic field provided by a solenoid-type magnet.
By rotating the cavity, the microwave magnetic field can also be perpendicular to a horizontal external field when a split-pair magnet is used.
Furthermore, a tiny metal cylinder allows for the adjustment of coupling.
This enables both continuous-wave (CW) and pulsed EPR experiments.
The coupling-varying ability has been demonstrated by reflection coefficient (S11) measurement, and CW and pulsed EPR experiments have been conducted.
The performance data indicates a prospect of wide applications of the cavity in the fields of physics, chemistry and biology.

\end{abstract}


\maketitle 

\section{Introduction}
High field electron paramagnetic resonance (EPR) spectroscopy is widely used  in many fields,
  such as biology,\cite{Mobius2005, Duss2014, Song2016} chemistry\cite{Chaudhuri2009, Roessler2018} and physics.\cite{Du2009, Flower2019}
It offers many advantages over traditional EPR spectroscopy, such as S-band and X-band EPR.
High field EPR provides higher spectral resolution, stronger orientational selectivity and less requirement on sample volume,
  and opens the path to understanding the fast motional dynamics of molecules.\cite{Mobius2009}

The microwave (mw) resonator is one of the most significant part of an EPR spectrometer, since it greatly influences the performance of the spectrometer.
At W-band, a single-mode cavity is usually the best choice for resonance because of
  its high microwave-power-to-magnetic-field conversion factor, relatively high filling factor and appropriate size.\cite{Webb2014, Rong2011}
These properties results in a high sensitivity and make the cavity easy to fabricate.
When the microwave is coupled into the resonant cavity,
  it generates an oscillating magnetic field $\bf{B_{1}}$, perpendicular to the external static magnetic field $\bf{B_{0}}$, and resonance absorption of the sample occurs.
In CW experiments, critical coupling between the cavity and the waveguide is desired for perfect impedance matching and maximized sensitivity.
Additionally, a modulation magnetic field $\bf{B_{m}}$ is applied in order to suppress the $1/f$ noise.
On the contrary, overcoupling is demanded in pulsed EPR for broad bandwidth, high sensitivity, and short deadtime.\cite{Pfenninger1995}


Magnets used in EPR experiments to provide external static magnetic field $\bf{B_{0}}$ fall into two types: solenoids twisting around cavities
  and split-pair magnets with cavities between the coils.
In most cases, a resonant cavity that is designed for split-pair magnets doesn't work in solenoids, and vise versa, which limits its applications.
A solenoid-type magnet providing a vertical $\bf{B_{0}}$ requires
  the cylindrical axis (the direction of $\bf{B_{1}}$) of the cavity to be in the horizontal plane,
  and the modulation coil parallel to the solenoid.
In the case of a split-pair magnet, however, the axis of the cavity is oriented vertically, and the axis of the modulation coil is in the same direction as the magnet.
Such configuration ensures $\bf{B_{1}}$ perpendicular to $\bf{B_{0}}$ and allows for convenient sample exchange from the top of the cavity.
Another obstacle for a universally compatible resonant cavity is the design for coupling.
When a solenoid is used, the coupling strength is usually adjusted by an iris\cite{Burghaus1992},
  or changing the relative position between the cavity and the waveguide\cite{Gromov1999,Annino2005}.
A novel method of adjusting the coupling using a metal sphere has been developed for cavities operating in split-pair magnets\cite{Savitsky2013}.
Nevertheless, each of these designs is valid for only one class of cases, e.g. the external magnetic field is provided by either solenoids or split-pair coils.

In this work, we report a novel design of resonant cavity.
This design is compatible with both solenoids and split-pair magnets.
The cavity operates on the T$_{011}$ mode and its cylindrical axis is in the horizontal plane.
The microwave magnetic field $\bf{B_{1}}$ is always perpendicular to the solenoids-provided external field, and by rotating the cavity,
  $\bf{B_{1}}$ can also be perpendicular to the external field provided by split-pair magnets.
Three modulation coils are integrated with the cavity.
They are selectively activated depending on the working conditions.
In both conditions, the coupling strength of the cavity is adjusted through a tiny metal cylinder.
The variable coupling mechanism can be achieved by remote control, and is compatible with the T$_{011}$ cavity in the horizontal field geometry.
A series of experiments are conducted to test the cavity.
The coupling-adjusting ability is verified and the performance of the cavity is demonstrated by CW and pulsed EPR experiments.

\section{Design}

\begin{figure}
  \centering
  \includegraphics[width = 0.5\textwidth]{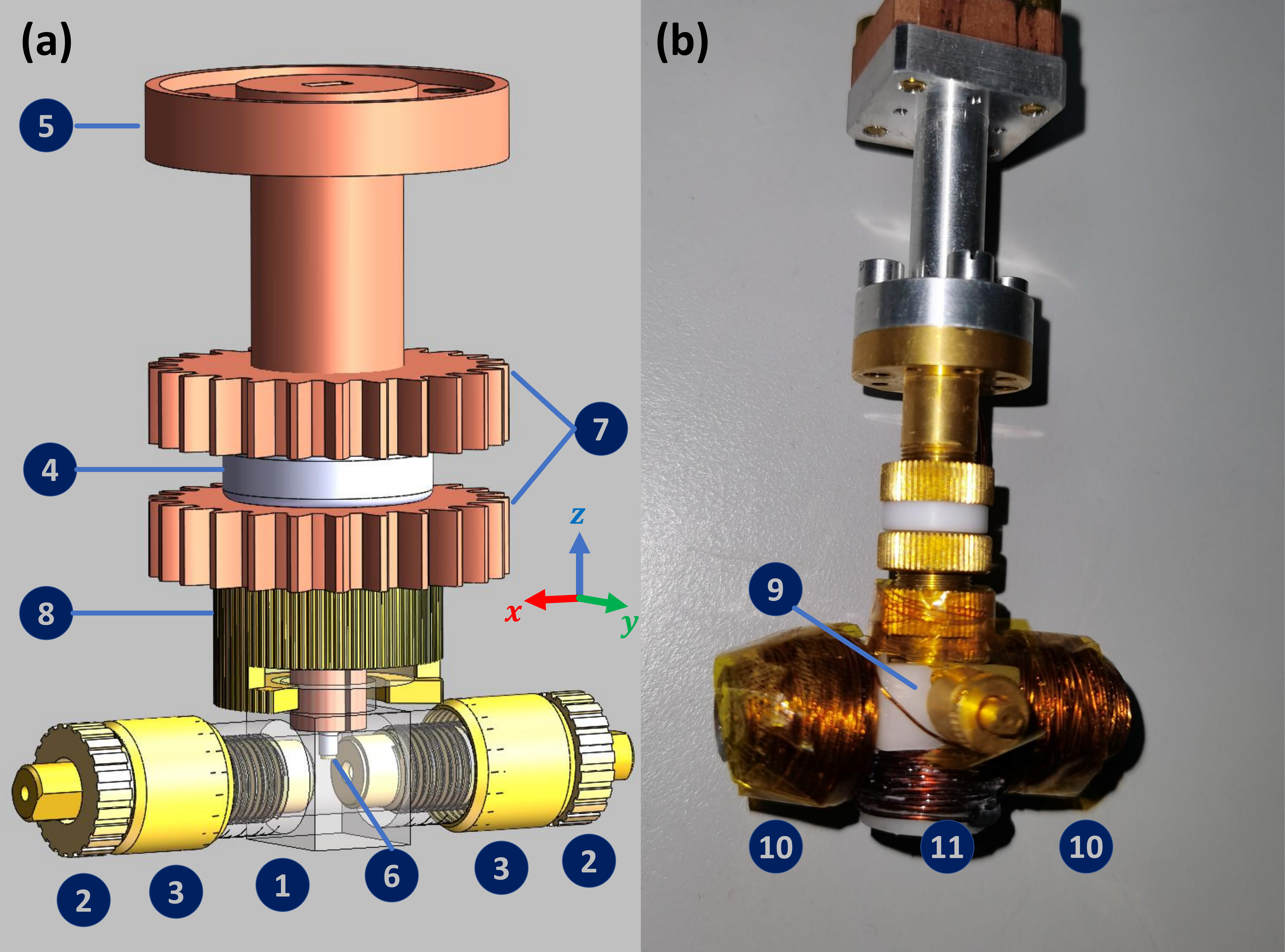}
 \caption
 {
  (a) Schematic diagram of the resonant cavity.
  (b) Photograph of the actual cavity.
  (1)The main body of the cylindrical cavity,
  (2)the tuning pistons,
  (3)sleeves of the pistons,
  (4)the Teflon plunger that holds the coupling adjusting cylinder,
  (5)the flange that connects the cavity to the waveguide,
  (6)the coupling adjusting cylinder,
  (7)the gears used for moving the Teflon plunger.
  (8)the hoop.
  (9)the Teflon of the modulation coils, and
  (10)(11)the modulation coils.
 }
  \label{FIG.1}%
\end{figure}

The structure of the cavity is shown in Fig.~\ref{FIG.1}.
It is a cylindrical brass cavity [Fig.~\ref{FIG.1}(a-1)] operating on the TE$_{011}$ mode, which is similar to that described before.\cite{Gromov1999}
The inner diameter of the cavity is 4.22 mm.
The inner surface of the cavity is plated with silver to improve the Q factor, and thus enhance the sensitivity.

The resonant frequency of the cavity has a variation range from 92.3 GHz to 95.6 GHz.
Delicate resonant frequency adjustment is realized by moving the pistons [Fig.~\ref{FIG.1}(a-2)] at the two ends of the cavity in and out,
  which changes the length of the cavity.
A micrometer-like structure, including the screw threads on the pistons and the main body of the cavity, together with the sleeves [Fig.~\ref{FIG.1}(a-3)], allows for
  subtle adjustments of the position of the pistons.

A Teflon plunger [Fig.~\ref{FIG.1}(a-4)] holding a
  tiny metal cylinder [Fig.~\ref{FIG.1}(a-6)] is inserted in the end of the copper flange [Fig.~\ref{FIG.1}(a-5)].
The flange is used for connecting the cavity to the rectangular waveguide tube, and is fastened to the cavity by a hoop [Fig.~\ref{FIG.1}(a-8)].
Variable coupling is achieved by the metal cylinder,
  which is inserted into the tail of the plunger and plugged into the 1.1-mm (inner diameter) coupling hole of the cavity.
Moving the cylinder up and down by two gears [Fig.~\ref{FIG.1}(a-7)]
  sandwiching the head of the plunger changes the coupling strength between the cavity and the waveguide.
This can be achieved automatically through a motor with a gear.
The coupling strength can be varied among undercoupling, critical coupling and overcoupling.
A typical loaded Q factor of the cavity at critical coupling is 2000.

\begin{figure}
  \centering
  \includegraphics[width = 0.5\textwidth]{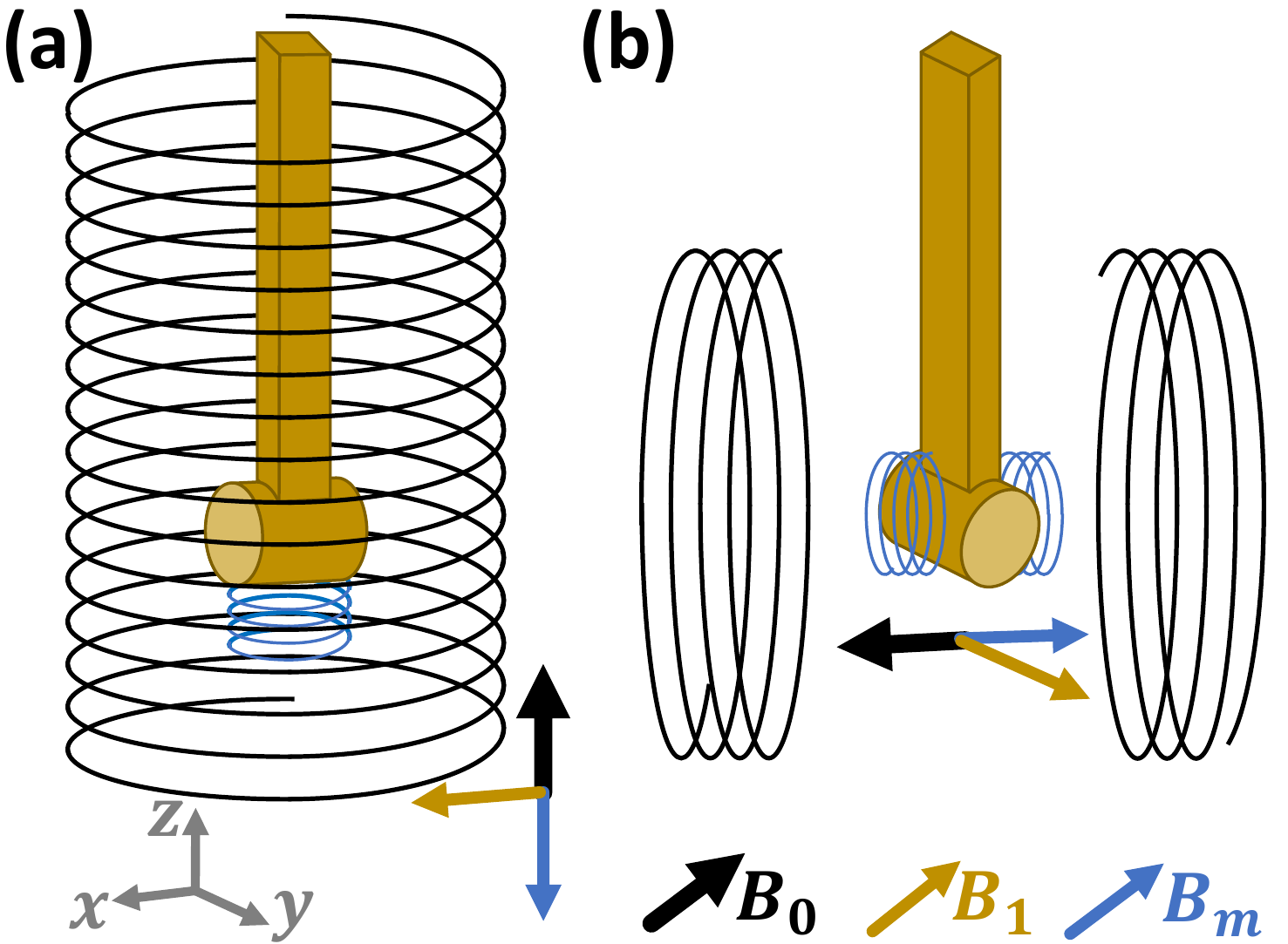}
 \caption
 {
  (a) Schematic diagram of the working mode with a solenoid-type magnet.
      The black spiral refers to the static magnet and the blue spiral refers to the modulation coils.
  (b) Schematic diagram of the working mode with a split-pair magnet.
      The black coils refer to the static magnet and the blue coils refer to the modulation coils.
      We can rotate the cavity around z axis to meet the working condition.
 }
  \label{FIG.2}
\end{figure}

Three coils [Fig.~\ref{FIG.1}(b-10,11)] wounding around the Teflon support [Fig.~\ref{FIG.1}(b-9)] provide the modulation fields.
For effective penetration of the modulation field, the modulation frequency $f_{m}$ is set to 6.25 kHz.

The direction of the microwave magnetic field $\bf{B_{1}}$ is along the cylindrical axis of the cavity, e.g. always in the xy-plane.
Therefore, it is perpendicular to the external magnetic field provided by solenoids, in the direction $\bf{\hat{z}}$, as shown in the Fig.~\ref{FIG.2}(a).
When the cavity operates in a solenoid, the coil below it is activated to generate a modulation magnetic field $\bf{B_{m}}$ that is parallel to $\bf{B_{0}}$.
The situation where the external magnetic field is provided by Helmholtz coils is shown in Fig.~\ref{FIG.2}(b),
  where the direction of $\bf{B_{0}}$ is along the x-axis.
The resonant cavity can be rotated around z-axis with the horizontal modulation coils enabled
  so that $\bf{B_{1}}$ is in the direction of $\bf{\hat{y}}$, perpendicular to $\bf{B_{0}}$,
  and $\bf{B_{m}}$ is in the same direction as the external field.

The sample area is up to 0.9 mm in diameter.
In this region, the magnetic field reaches the strongest and the electric field is almost zero.
During experiments, samples in a sample tube are loaded into the cavity from the hole at the axis of the piston.
The microwave-power-to-magnetic-field conversion factor is $4.7~ \mathsf{Gauss}/\sqrt{\mathsf{W}}$ obtained by Rabi oscillation experiment described in the following.
Such relatively high conversion factor allows high signal-to-noise ratio even at low sample concentrations and short $\pi$ and $\pi/2$ pulses.

\section{Experimental Setup}
\begin{figure}
  \centering
  \includegraphics[width =0.5\textwidth]{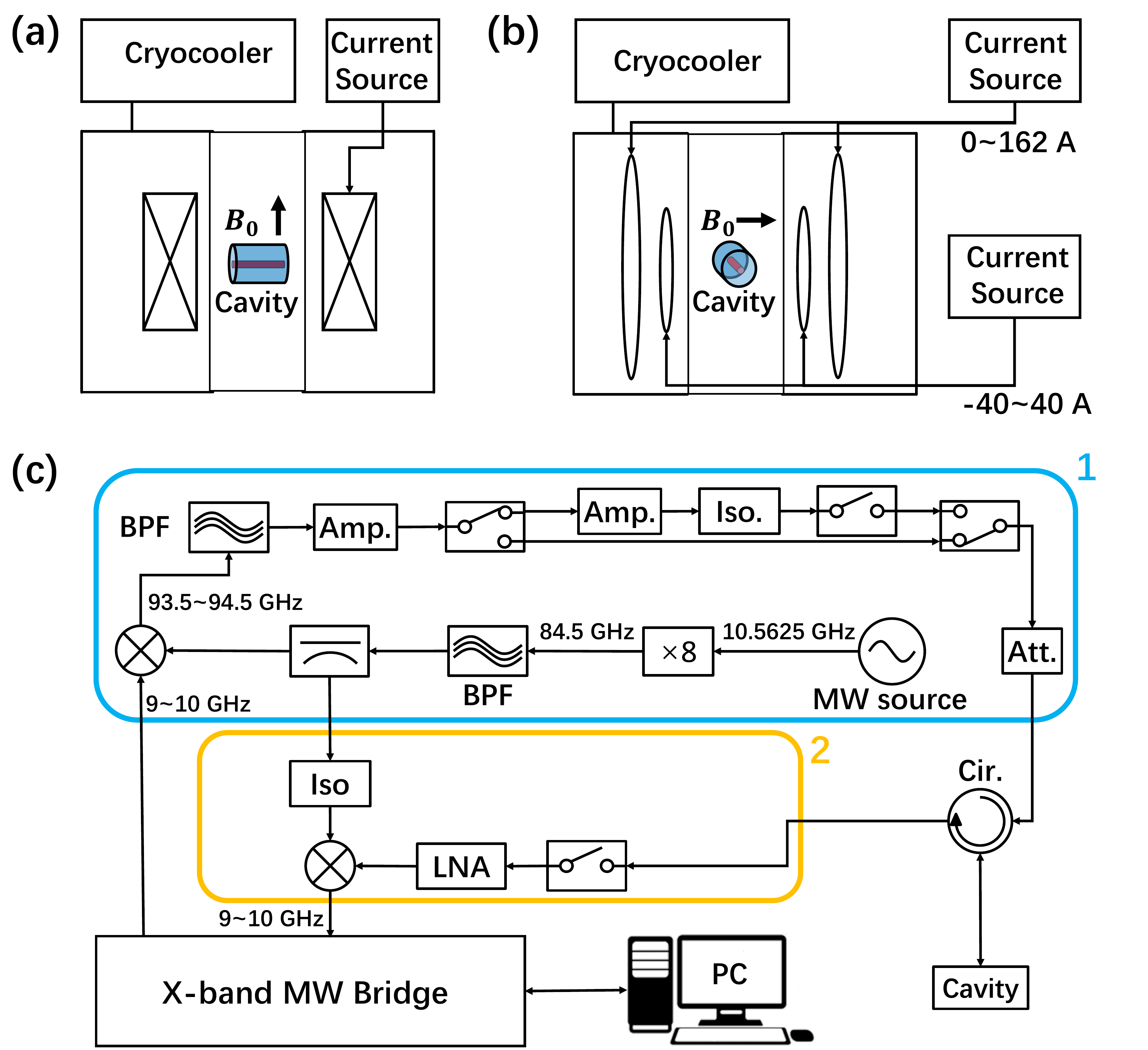}
  \caption
  {
    (a) Block diagram of the W-band EPR spectrometer equipped with a solenoid-type magnet.
    (b) Block diagram of the W-band EPR spectrometer equipped with a split-pair magnet.
    (c) The detailed information about the W-band EPR spectrometer. The Helmholtz coils are cooled down to below 3 K to maintain superconductive.
    The W-band mw bridge is built based on an X-band mw bridge we developed before.
    The X-band mw bridge outputs and processes a 9$\sim$10-GHz IF mw signal.
    Box 1 stands for the transmitter part.
    An 84.5-GHz microwave is generated and mixed with the IF signal, forming a 93.5$\sim$94.5-GHz W-band microwave.
    The mw bridge is able to operate in both CW and pulsed modes.
    Switching between the two modes is realized by two switches.
    Box 2 stands for the receiver part.
    The EPR signal is amplified by a low-noise amplifier (LNA) and mixed with the 84.5-GHz reference signal.
    The output IF signal is fed into the X-band mw bridge and further processed.
  }
  \label{FIG.3}%
\end{figure}

For testing of this novel resonant cavity, a W-band EPR spectrometer named EPR-W900, which has been developed by University of Science and Technology of China and Chinainstru $\&$ Quantumtech (Hefei) Co.,Ltd, is utilized.
As shown in Fig.~\ref{FIG.3}, beside the cavity, the spectrometer includes two magnets and a microwave bridge.

\subsection{Magnets}
Two types of magnets have been utilized in our experiments to show that the reported cavity can be used with these magnets. One is a solenoid-type magnet, which is produced by Cryomagnetics.inc. The other is a split-pair magnet. For the split-pair magnet, two pairs of Helmholtz coils are used to provide the external field.
All of them are cooled down to below 3 K using a Gifford-McMahon cryocooler and compressor so that the coils are superconductive.
The compressor is cooled by water.
The current passing through the main coils can be at most 162 A, creating a static magnetic field up to 6 T.\
High electrical inductance of the main coils limits the rate of change of current ($\leq 0.02$ A/s).
For saving time when conducting narrow filed-swept experiments, a second pair of field-sweeping coils is used.
The field-sweeping coils carry currents from -40 to 40 A, corresponding to the magnetic field from -0.1 T to 0.1 T,
 and the maximum rate of change of current on it is 1.6 A/s.

\subsection{Microwave bridge}
The W-band microwave bridge is based on a commercial X-band microwave bridge we developed before,\cite{Shi2018}
  which outputs and receives a 9$\sim$10 GHz intermediate frequency (IF) mw signal.
A mw signal of 10.5625 GHz is generated by a mw source and octupled and filtered, forming an 84.5-GHz microwave.
It is used to convert an X-band signal to a W-band one or vise versa, enabling W-band excitations and readouts.
One arm of the 84.5-GHz mw is up-converted with an IF signal of 9$\sim$10 GHz to form a W-band signal of 93.5$\sim$94.5 GHz.
Then the W-band mw signal is filtered and amplified by a band-pass filter (BPF) and an amplifier (Amp.) before the mode-selecting switch.
In the CW mode, the mw signal straightly passes the switch and a variable attenuator (Att.),
  then is fed into the cavity through a circulator (Cir.).
In the pulsed mode, a cascade of
  an amplifier, an isolator and a switch driven by high-speed square waves is inserted between the amplifier and the variable attenuator.
Switching between the two modes is controlled by the computer.
Another arm of the 84.5-GHz mw is mixed with the EPR signal amplified by a low-noise amplifier (LNA).
The output of the mixer is an IF signal of 9$\sim$10 GHz, which is then processed by the X-band mw bridge.

\section{Performance}
The ability of the cavity to change the coupling strength is demonstrated by reflection coefficient (S11) measurement.
For testing the EPR performance of the cavity, both continuous-wave experiments and pulsed experiments are carried out.

\subsection{Coupling adjusting ability}
\begin{figure}[!h]
  \centering
  \includegraphics[width = 0.5\textwidth]{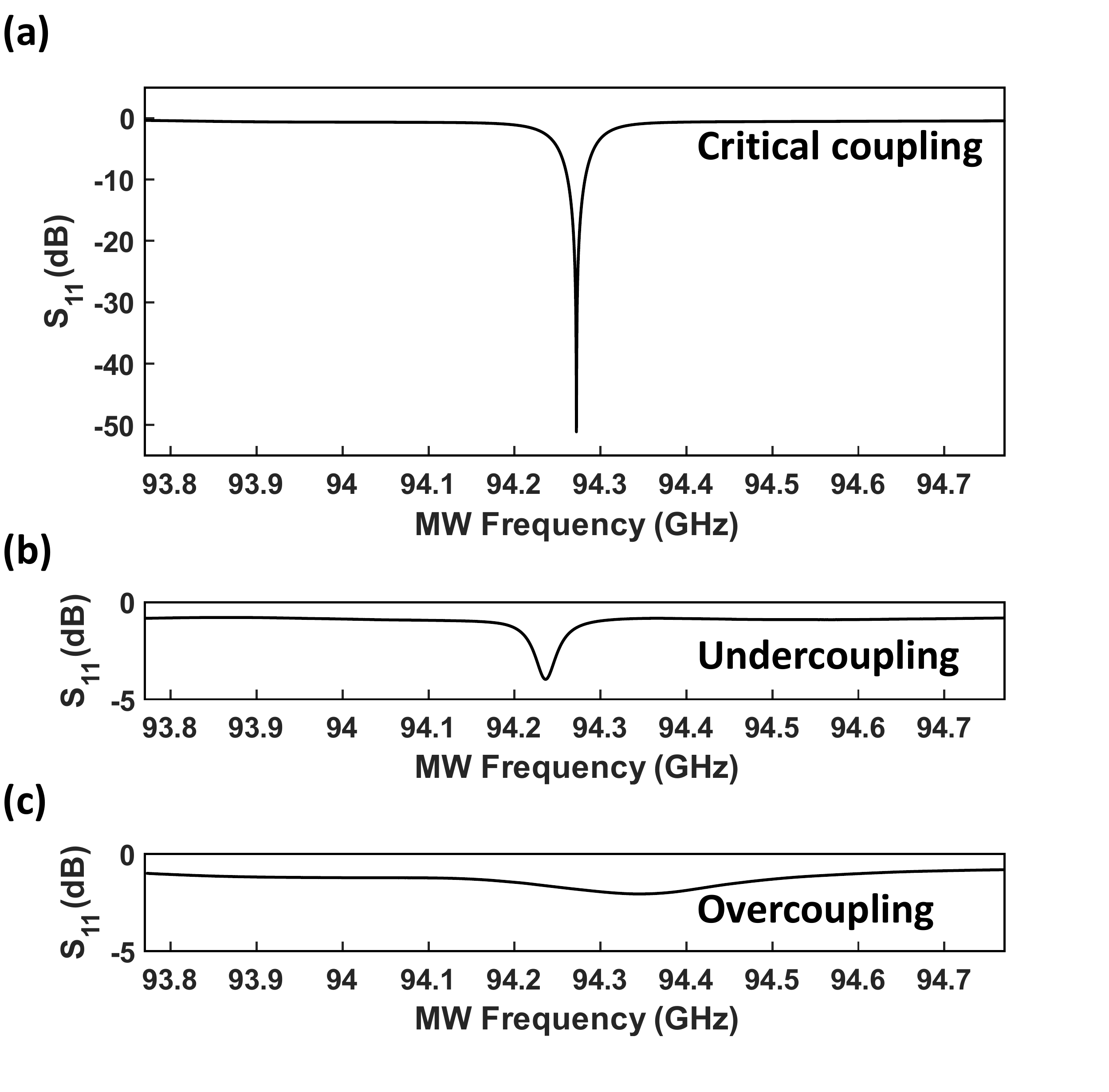}
  \caption
  {
Reflection coefficient S$_{11}$ versus frequencies in different coupling conditions.
    (a) Reflection coefficient at critical coupling.
    (b) Reflection coefficient at undercoupling.
    (c) Reflection coefficient at overcoupling.
  }
  \label{FIG.4}
\end{figure}

We test the reflection coefficient ($S_{11}$) at different coupling strengths of the empty resonant cavity.
The mw frequency is swept from 93.77 GHz to 94.77 GHz.
Power of the reflected microwave is measured and plotted in Fig.~\ref{FIG.4}.
Reflection coefficient measurement at critical coupling is shown in Fig.~\ref{FIG.4}(a).
The extremely sharp and deep absorption peak is a symbol of high frequency selectivity and strong absorption,
  which is preferred in CW EPR experiments.
Fig.~\ref{FIG.4}(b) shows the undercoupling $S_{11}$ curve.
The absorption peak is narrow but shallow, indicating relatively high frequency selectivity but weak absorption.
The condition of overcoupling is shown in Fig.~\ref{FIG.4}(c).
The peak is very shallow and broad.
This is because in overcoupling status the cavity is able to absorb microwaves in a wide range of frequencies,
  resulting in a broad bandwidth.
Pulsed EPR experiments are carried out in this range.

\subsection{CW EPR experiment}
\begin{figure}
  \centering
  \includegraphics[width = 0.5\textwidth]{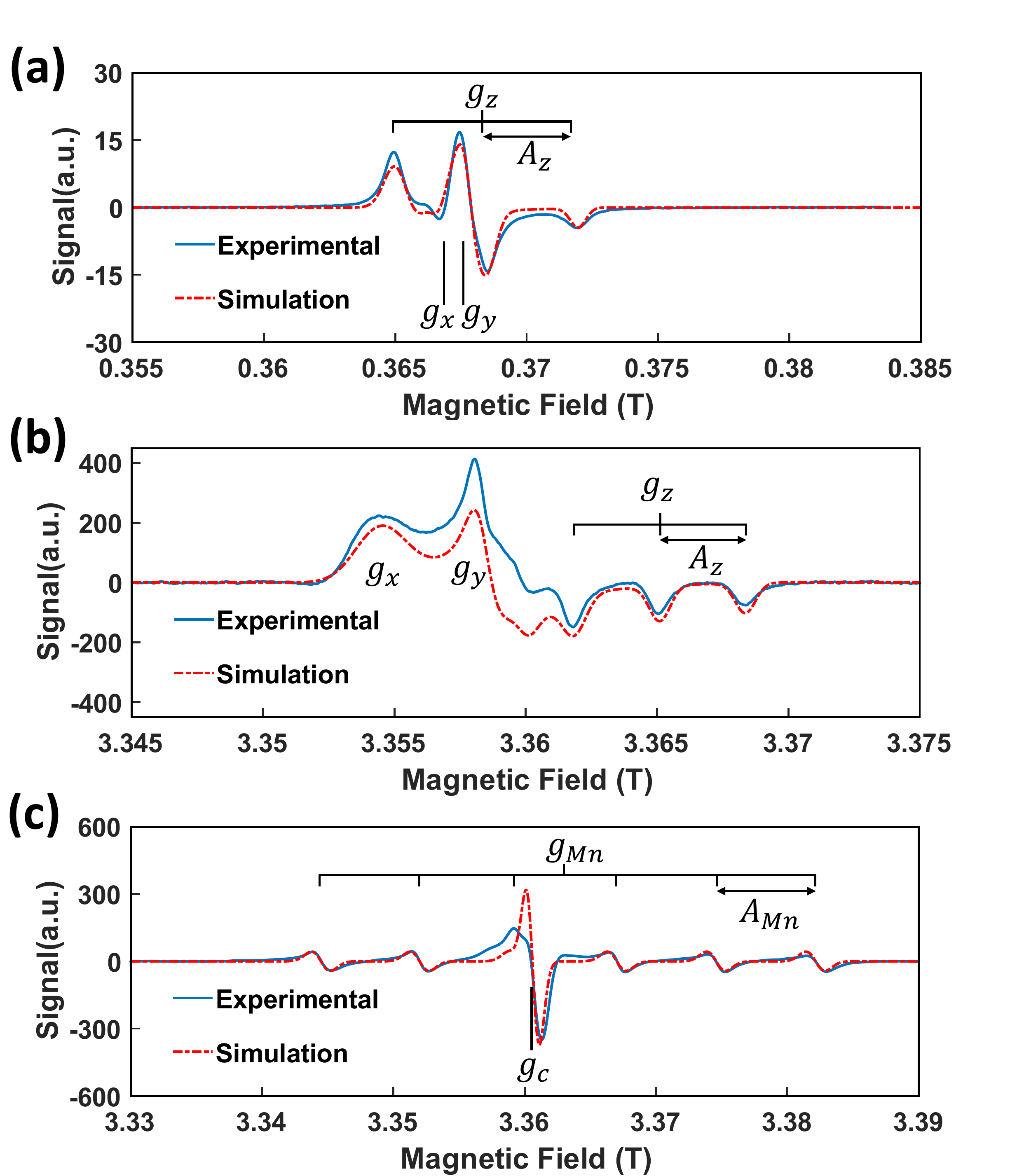}
  \caption
  {
    Experimental and simulated CW EPR spectra of  starch at room temperature with modulation amplitude of 2 Gauss. Blue lines are experimental data and the red dashed lines are the simulations.
    (a) The X-band EPR spectrum of starch. The frequency of the microwave is 9.82 GHz and the power is 1.0mW. The modulation frequency is 100 kHz.
    (b) The W-band EPR spectrum of starch with a split-pair magnet.
        The mw frequency is 94.2 GHz with an input power of 0.63mW.
        The modulation frequency is 6.25 kHz.
    (c) The W-band EPR spectrum of lignite with a solenoid-type magnet. The mw frequency is 94.1 GHz with an input power of 0.63 mW.
        The modulation frequency is 6.25 kHz.
  }
  \label{FIG.5}
\end{figure}

Continuous-wave EPR is utilized for analyzing the properties and structures of materials.\cite{Goswami2015,Kempe2010}
Field-swept EPR experiments at room temperature with starch and lignite powder were carried out.
Nitrogen oxide radicals are tagged to starch molecules by chemical synthesis.
For the sample being starch, both X-band and W-band room temperature powder EPR spectra have been recorded as shown in Fig.~\ref{FIG.5}(a) and (b).
The solid blue lines stand for the experimental result and the dashed red lines stand for the simulation. For the W-band EPR experiment in Fig. ~\ref{FIG.5} (b), a split-pair magnet has been utilized.
The input power of the microwave is 0.63 mW.
The external field $\bf{B_{0}}$ is swept from 3.325 T to 3.355 T.
At W-band, the experimental spectrum of starch fits well with the simulated one, but is totally different from that at X-band.
The characteristic shape of the spectrum is due to the anisotropy of g factor and hyperfine splitting.\cite{Mobius2009}
The $g_{zz}$ peak splits into three because of the $A_{zz}$ hyperfine-tensor component.
They are corresponding to the three peaks observed in Fig.~\ref{FIG.5}(a).
The $g_{xx}$ and $g_{yy}$ peaks, which are hidden at X-band, are resolved clearly at W-band.
Fig.~\ref{FIG.5}(c) shows the experimental result at W-band using the cavity with  a solenoid-type magnet, when the sample is lignite powder.
The power of the microwave is set to 0.6 mW and the frequency of the microwave is 94.1195 GHz.
The frequency of the modulation field is 6.25 kHz, and the amplitude of the modulation field is 2 Gauss.
Our experiments show that the cavity suits with both types of magnets.

\subsection{Pulsed EPR experiment}
Pulsed EPR technology has a wide range of applications in fields such as quantum information.\cite{Morton2005}
It also helps to understand the spin-spin and spin-environment interactions, which is highly concerned in chemistry and biology.\cite{Schweiger2001}
For testing the performance of the cavity in pulsed experiments, Rabi oscillation and relaxation-time experiment are conducted with lignite.

\begin{figure}
  \centering
  \includegraphics[width = 0.5\textwidth]{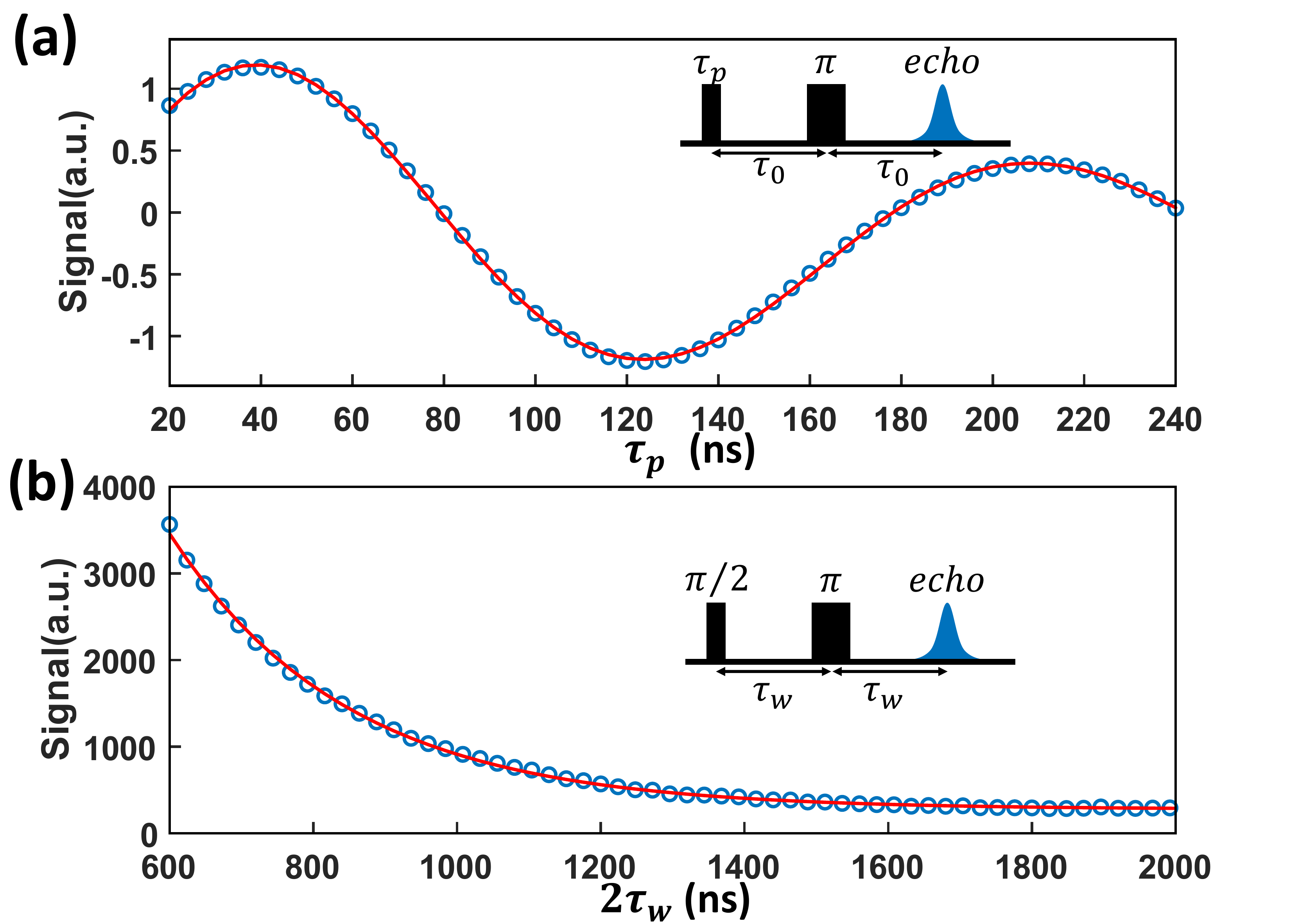}
  \caption
  {
    Pulsed EPR experiments with lignite.
    The inset in each subplot shows the pulse sequence for each experiment.
    The microwave has the frequency of 93.7343 GHz and input power of 200 mW. The pulse lengths of the $\pi/2$ pulse and  $\pi$ pulse are 42 and 85 ns, respectively.
    (a) The Rabi oscillation experiment.
        The duration of the first pulse $\tau_{p}$ is swept from 20 ns to 240 ns in the 4-ns step.
        The interval between two pulses $\tau_{0}$ is set to 300 ns.
    (b) Experimental data of measurement of the transverse relaxation time.
        The interval between two pulses $\tau_{w}$ varies in the 24-ns step with an initial value of 300 ns.
        The fitting gives $T_{2} = 248.3 \pm 4.0\ ns$.
  }
  \label{FIG.6}%
\end{figure}

For Rabi oscillation experiment, the microwave pulse sequence is shown as the inset of Fig.~\ref{FIG.6}(a).
The frequency of the microwave is 93.7343 GHz and the peak microwave power is 200 mW.
The direction of the external field $\bf{B_{0}}$ is defined as z-axis and that of $\bf{B_{1}}$ as y-axis.
Firstly a pulse with duration $\tau_{p}$ is applied to rotate Bloch vectors of the spins around the x-axis in the frame
  that is rotating with $\bf{B_{1}}$.
Bloch vectors rotate around the external field at Larmor frequency $\omega_{L} = \gamma_{e} B_{0}$.
Since the external field is not perfectly uniform, the vectors have slightly different angular speeds and diffuse when rotating.
After a period of evolution time $\tau_{0}$, a $\pi$ pulse is applied so the Bloch vectors rotate 180 degrees around the x-axis and begin to refocus.
After another $\tau_{0}$ the effect of diffusion is almost cancelled and the Bloch vectors gather in one place, forming a spin echo.
Here we set $\tau_{0}$ as 300 ns.
As Fig.~\ref{FIG.6}(a) shows,
  different lengths of the first pulse are corresponding to different polarizations, so the amplitude of the echo signal oscillates with the pulse length.
The pulse length of the $\pi/2$ ($\pi$) pulse is measured to be 42 (85) ns. The amplitude of the B1 field is 2.1 Gauss while the power of the microwave pulses is 200 mW. Thus the microwave-power-to-magnetic-field conversion factor is  $4.7~\mathsf{Gauss}/\sqrt{\mathsf{W}}$.

To measure the transverse (spin-spin) relaxation time $T_{2}$,
  we set the first mw pulse as the $\pi/2$ pulse and vary the evolution time $\tau_{w}$ in the Hahn echo sequence.
Since the spins are not isolated but interact with each other,
  the Bloch vectors in the xy-plane continue to shorten, and as a result, the refocused spin echo exponentially decays over time.
The echo signal can be fitted by
\begin{equation}
  S = A_{1}\exp(-t/T_{2})+A_{2},
\end{equation}
where $S$ is the signal intensity, $T_{2}$ is the spin-spin relaxation time, $A_{1}$ and $A_{2}$ are undefined constants.
According to the experimental data shown in Fig.~\ref{FIG.6}(b), the $T_{2}$ is 248.3 $\pm$ 4.0 ns.


\section{Conclusion}
To conclude, we have designed a new type of universally compatible resonant cavity for W-band EPR.
The cavity is compatible with both types of magnets that are commonly used in EPR spectrometers.
It is installed in a W-band EPR spectrometer and tested with two types of magnets.
The range of the coupling strength proves to cover under-, critical and over-coupling.
The results of field-swept CW experiments show that our cavity suits with both types of magnets.
By pulsed EPR experiments, it is demonstrated that the cavity is qualified for quantum state manipulation and dephaseing-time measurements. The microwave-power-to-magnetic-field conversion factor up to $4.7\ \mathsf{Gauss}/\sqrt{\mathsf{W}}$ is also achieved.

\begin{acknowledgments}
This work was supported by the Chinese Academy of Sciences (Grants No. XDC07000000, No. GJJSTD20200001). X. R. thanks the Youth Innovation Promotion Association of Chinese Academy of Sciences for the support.
\end{acknowledgments}

\end{document}